\documentclass{article}

    \usepackage{neurips_2020}

\usepackage[utf8]{inputenc} % allow utf-8 input
\usepackage[T1]{fontenc}    % use 8-bit T1 fonts
\usepackage{hyperref}       % hyperlinks
\usepackage{url}            % simple URL typesetting
\usepackage{booktabs}       % professional-quality tables
\usepackage{amsfonts}       % blackboard math symbols
\usepackage{nicefrac}       % compact symbols for 1/2, etc.
\usepackage{microtype}      % microtypography
\usepackage{tikz}
\usepackage[figuresright]{rotating}
\usepackage{graphicx}
\usepackage{amsfonts,amssymb,amsmath}
\usepackage{boxedminipage}
\usepackage{makeidx}
\usepackage{multicol}

\usepackage{fancyhdr}

\title{Sequence-based Machine Learning Models in Jet Physics}

\author{%
  Rafael Teixeira de Lima\\
  SLAC National Accelerator Laboratory\\
  \texttt{rafaeltl@slac.stanford.edu}
}

\begin{document}

\maketitle

\begin{abstract}
  Sequence-based modeling broadly refers to algorithms that act on data that is represented as an ordered set of input elements. In particular, Machine Learning algorithms with sequences as inputs have seen successfull applications to important problems, such as Natural Language Processing (NLP) and speech signal modeling. %Due to the nature of these applications, such models are optimized for situations in which the input sequences have different lengths. 
  The usage this class of models in collider physics leverages their ability to act on data with variable sequence lengths, such as constituents inside a jet. 
  In this document, we explore the application of Recurrent Neural Networks (RNNs) and other sequence-based neural network architectures to classify jets, regress jet-related quantities and to build a physics-inspired jet representation, in connection to jet clustering algorithms. In addition, alternatives to sequential data representations are briefly discussed.
\end{abstract}

\tableofcontents
\newpage
\section{Introduction}

The area of \textit{Sequence-Based Learning} in Machine Learning deals with the concepts and algorithms used to learn from data represented as an ordered set (sequence) of objects, each with its set of characteristics (features), in which positional information of each object (context) is important. 
The idea of contextual information as being important for the algorithms is fundamental, since it can encode correlations between objects along the sequence. 
One of the main applications of this class of models is in natural language processing (NLP). 
In these cases, the sequence is often built from words in a sentence and the algorithm must learn from it. 
How the learning occurs and what is learned will depend on the application. 
To perform a translation task (neural machine translation), for example, the algorithm must output another sequence. 
In other instances, a summary semantic information needs to be obtained, such as when the algorithm needs to classify a certain sentence as positive or negative in an on-line product review, for example. 

\begin{figure}[h]
\begin{center}
\begin{tikzpicture}
\draw[black,thick] (-2,0) circle[radius=0.5] ;
\draw[black,thick] (0,0) circle[radius=0.5] ;
\draw[black,thick] (2,0) circle[radius=0.5] ;
\draw[->, thick] (-3.5,0) -- (-2.5,0);
\draw[->, thick] (-1.5,0) -- (-0.5,0);
\draw[->, thick] (0.5,0) -- (1.5,0);
\draw[->, thick] (2.5,0) -- (3.5,0);
\draw (-2,0) node {$\mathbf{x}^{t-1}$};
\draw (0,0) node {$\mathbf{x}^{t}$};
\draw (2,0) node {$\mathbf{x}^{t+1}$};
\draw (-3,0.25) node {$f$};
\draw (-1,0.25) node {$f$};
\draw (1,0.25) node {$f$};
\draw (3,0.25) node {$f$};
\end{tikzpicture}
\caption{Scheme of a sequence-based algorithm acting on a sequence of inputs $\mathbf{x}^{t}$.}
\label{fig:seq-mod}
\end{center}
\end{figure}
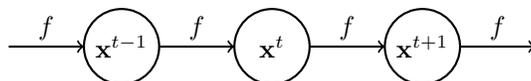

In more mathematical terms, sequence-based models aim to perform operations $f$ on a sequence of inputs $\{\mathbf{x}^{t}\}$, where each entry $\mathbf{x}^{t}$ is a vector of features, and $t$ is a position in the ordered sequence with a length $T$, as shown in the scheme presented in Fig. \ref{fig:seq-mod}. 
In particle physics terms, the sequence can represent an ordered set of tracks that constitutes a jet, for example, while the entry $\mathbf{x}^{t}$ represents the kinematics of the track in the position $t$ in that sequence. 
An algorithm acting on the sequence $J = \{\mathbf{x}^{t}\}$ can then be used to learn information about that jet, such as its flavor or charge (as will be discussed in the following sections).

In general, different sequences being utilized for the algorithm definition will have different lengths, just as jets can have different number of tracks. 
If all sequences have fixed length, they can be collapsed into a single feature vector and simpler algorithms, such as densely-connected NNs, can be used. 
However, even for fixed-length sequence problems, sequence-based models can outperform simpler models by exploiting the ordered nature of the input data. 

% \begin{figure}
% \begin{center}
% \begin{tikzpicture}

% \draw[black,thick] (-2,0) circle[radius=0.5] ;
% \draw[black,thick] (0,0) circle[radius=0.5] ;
% \draw[black,thick] (2,0) circle[radius=0.5] ;

% \draw (-2,0) node {$\mathbf{h}^{t-1}$};
% \draw (0,0) node {$\mathbf{h}^{t}$};
% \draw (2,0) node {$\mathbf{h}^{t+1}$};

% \draw[->, thick] (-3.5,0) -- (-2.5,0);
% \draw[->, thick] (-1.5,0) -- (-0.5,0);
% \draw[->, thick] (0.5,0) -- (1.5,0);
% \draw[->, thick] (2.5,0) -- (3.5,0);

% \draw[->, thick] (-2,-1.5) -- (-2,-0.5);
% \draw[->, thick] (0,-1.5) -- (0,-0.5);
% \draw[->, thick] (2,-1.5) -- (2,-0.5);

% \draw[black,thick] (-2,-2) circle[radius=0.5] ;
% \draw[black,thick] (0,-2) circle[radius=0.5] ;
% \draw[black,thick] (2,-2) circle[radius=0.5] ;

% \draw (-2,-2) node {$\mathbf{x}^{t-1}$};
% \draw (0,-2) node {$\mathbf{x}^{t}$};
% \draw (2,-2) node {$\mathbf{x}^{t+1}$};

% \draw (-3,0.25) node {$f$};
% \draw (-1,0.25) node {$f$};
% \draw (1,0.25) node {$f$};
% \draw (3,0.25) node {$f$};
% \end{tikzpicture}
% \caption{Scheme of a sequence-based algorithm acting on a sequence of inputs $\mathbf{x}^{t}$.}
% \label{fig:rnn}
% \end{center}
% \end{figure}

A key feature of sequence-based models is the ability to share parameters in different parts of the same model, i.e., the operation $f$, which is learned, is applied to every step in the sequence. 
With parameter sharing, these models are able to generalize to sequences of different lengths, using the same set of parameters throughout the input elements. 
If different parameters were to be learned individually, the desired generalizability would not be achieavable, and the model would behave similar to a densely-connected network.

\subsection{Recurrent neural networks}
\label{ssec:rnn}

A recurrent neural network (RNN) is a neural network implementation of the concepts described in the previous section. 
Densely-connected neural networks map an input vector of features $\mathbf{x}$ to an output vector $\mathbf{o}$. In contrast, RNNs map a sequence of inputs ${\mathbf{x}}$ into an output, which can be a vector or a sequence as well. 
This difference can be achieved in many different ways, but RNN architectures generally present cyclical connections between units in the same or different layers. 

These interconnections between units are sequential, in the sense that each unit's hidden state is obtained by a combination of the previous unit's hidden state and the input from that step. 
This means that instead of a unit's hidden state be given by $h = f(\mathbf{x}; \theta)$, it will be given by $h^t = f(h^{t-1}, \mathbf{x}^{t}; \theta)$. On the particular case in which $f$ is given by a hyperbolic tangent function, for example, the RNN can be represented as 
\begin{equation}
h^{t} = \tanh{ \left( \mathbf{W}^{\top} \mathbf{x}^{t} + \mathbf{V}^{\top} h^{t-1} + b \right) },
\label{eq:rnn}
\end{equation}
instead of a simple densely-connected unit $h = \tanh{ \left( \mathbf{W}^{\top} \mathbf{x} + b \right) }$, where $\mathbf{W}$, $\mathbf{T}$, and $b$ represent learnable weights and biases.

Notice that in eq.~\ref{eq:rnn} the weights on the operations ($\theta$, or $\mathbf{W}$, $\mathbf{V}$, and $b$ explicitly) do not depend on the time step, or the sequence element position, $t$, explicitly showcasing the parameter sharing feature of the RNN. 
Similarly to MLPs functioning as universal approximators, large enough RNNs have been shown to universaly approximate any measurable sequence-to-sequence maps~\cite{HAMMER2000107}.

In general, RNN architectures add other features on top of the recurrent layer format described above. 
For example, in tasks where the algorithm needs to output another sequence, each individual hidden state in the recurrent layer might be read out into a densely connected network. 
In contrast, the cases where only a single output is read at the end of the sequential layer (also known as time-unfolded RNNs) are used to extract a summary information of the input sequence.
Even though the presence of cycles in these architectures could potentially complicate the process of updating the network parameters during the optimization step, backpropagation can still be applied to the unrolled computation graph\footnote{The application of backpropagation on the time-series unrolled RNN is known as \textit{backpropagation through time}.}, thus no specialized algorithms are necessary.

A simple but powerful extension of standard RNN architectures are Bidirectional Recurrent Neural Networks (BRNNs)~\cite{SchusterPaliwal1997}. 
For certain sequence-based modeling applications, knowledge about backward-in-time context might be as important as forward-in-time, only the latter of which is exploited in standard RNNs. 
BRNNs manage to extend that context information by adding a second recurrent layer to the network architecture which processes the sequence in reversed order. 
Both the forward and backward RNNs are then connected to the same output layer, providing a combined representation. 

An idea related to RNN architectures which is also used in time-series and signal processing analyses are 1D convolutional layers. 
In these architectures, the convolution operation kernal acts on neighbouring time steps, and outputs a new sequence based on its inputs. 
These operations act on fixed length sequences, where empty entries can be masked - similarly to 2D CNNs acting on sparse images with a fixed pixel grid. 
Parameter sharing is also an important feature here, exploited by the use of a single convolution operation across different time steps.
One drawback of this method is its limited sensitivity to long-term dependencies, only exploring correlations across close neighbours, as defined by the kernel length.

\subsection{Long-Term Dependencies, LSTMs and GRUs}

When dealing with large sequences, one expected behavior of RNNs is to learn how correlated certain entries are, regardless of how far appart they appear in the input sequence. In practical tearms, this implies that information from early entries in the sequence must be encoded in how the network learns about latter entries (long-term dependencies). 
Unfortunately, in the learning steps, it is common for gradients propagated through many steps to either quickly approach zero or increase - both undesirable features in terms of optimization. 
This issue is referred in the literature as the \textit{vanishing or exploding gradient problem} (e.g. in~\cite{Pascanu2012}). 
Vanishing gradients, for example, can lead to long-term dependencies being given less importance through the sequence compared to short-term ones. 

This issue can be understood by imagining a simplified recurrent structure as a linear transformation between hidden states $h_t = \mathbf{V}^{\top}h_{t-1}$. This operation will be therefore performed $T$ times when moving from the time-step 0 to the last sequence entry. 
This shows that the learnable parameters in $\mathbf{V}$ will be raised to the power of $T$, which means that weights less than 1 will tend to approximate 0 at later steps, while weights larger than 1 will quickly grow.
Most modern RNN architectures solve this problem with the usage of Long Short-Term Memory~\cite{HochreiterSchmidhuber1997} (LSTM) units or Gated Recurrent Units~\cite{Cho2014} (GRUs).

LSTM units mitigate the vanishing gradient problem by introducing a new path in the recurrent loop where the information coming from each sequence entry can flow for long durations, possibly without interference from subsequent hidden states. 
This path is dynamically gated through learnable parameters, which means that the importance of long-term dependencies is optimized with the rest of the network parameters. 
In particular, if the activation function pertaining to a gate remains close to 0, the information from the previous time-step will not be propagated throughout the sequence. However, the LSTM will still use that time-step information to update its current hidden state. 
This optimizable gated structure ensures that both long-term and short-term contributions to the gradient are taken into account. 

\begin{figure}[ht]
    \centering
    \includegraphics[width=0.9\textwidth]{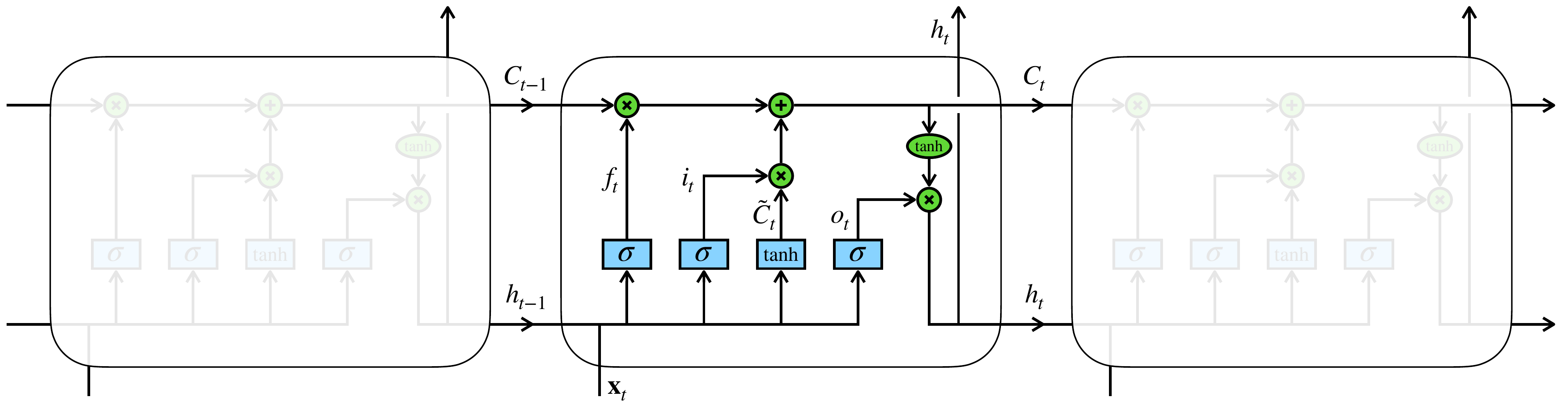}
    \caption{Schematic structure of a LSTM recurrent layer. Two outputs for the unit's hidden state $h_t$ are shown, representing the case in which the LSTM layer outputs another sequence. Image adapted from~\cite{colah_lstm}.}
    \label{fig:lstm_arch}
\end{figure}

A schemematic view of three sequential LSTM units is shown in Figure \ref{fig:lstm_arch}. 
The LSTM receives at the time step $t$ both the hidden state of the previous time step ($h_{t-1}$), which is concatenated with the feature vector $\mathbf{x}_t$, and an extra input (the cell state, $C_{t-1}$) which is regulated by a forget gate ($f_t$). 
The forget gate can be a neural network itself, with a sigmoid output which is shared across all units, ensuring that the LSTM actively learns how much long-term correlations should be propagated in the sequence. 
Next, the cell state is updated with information learned from the previous hidden state and $\mathbf{x}_t$ with a dedicated neural network, generating an intermediate cell state $\tilde{C}_t$. 
The impact of $\tilde{C}_t$ in the final cell state is regulated by another neural network, $i_t$. 
After the update, $C_{t}$ is propagated to the next time-step. 
The final hidden state at this time step is derived with information learned from the previous hidden state and $\mathbf{x}_t$ through a network ($o_t$), and also $\tilde{C}_t$.

GRUs work with a similar gated structure - however, the same forget gate that decides on the propagation of the previous cell state also determines, with an inverse importance, how the current cell state should be updated with $h_{t-1}$ and $\mathbf{x}_t$. 
With the LSTM nomenclature used above, the forget gate and the cell state update gate would be given by $f_t = u_t$ and $i_t = (1-u_t)$, respectively, where $u_t$ is called the update gate.
GRUs comparatively uses less trainable parameters than LSTMs, which can be benefitial for smaller datasets; its less flexible architecture can potentially be detrimental in more complex applications - however, the studies to be shown below that compared GRUs and LSTMs generally see similar performances.

\section{Applications of RNNs to Jet Physics}

Representing reconstructed jets in collider experiments as images  for classification and other machine learning based tasks, has been a successful avenue of research for years now.% ({\color{red} see Section XYZ}). 
There are, however, a few features related to this choice of representation that can make the process of training a computer vision-based algorithm difficult when compared, for example, to the training of a simple densely-connected neural network based on engineered features. 

Jet images can be very sparse, i.e., containing few populated pixels\footnote{Initial computer vision-based jet images studies have reported $5 \sim 10\%$ of activated pixels on average for 25 x 25 pixels images from signal and background-type jets \cite{OliveiraJetImages}.}, making the identification of features on individual jets a complicated task, even if identifying these features on their averaged images might be very easy. 
Pre-processing steps can be applied to reduce sparseness, for example, increasing the coarseness the image further by combining adjacent pixels. 
This procedure, however, effectively reduces the spatial information contained in the input image, penalizing the algorithm's performance.

Another issue arising in the pre-processing step is finding a unique geometrical representation for these images that reflects the expected symmetries of the problem. 
In general, geometrical transformations (e.g., rotations, translations and reflections) are used, aligning pre-defined axes based on the jets' spatial energy distributions - but these definitions can be very task specific and not well generalizable. 
This is specially true when the pattern of energy deposition inside these jets displays a different number of core clusters (prongness) - for example, when comparing quark-initiated jets, jets from a collimated hadronic W boson decay, and jets from a collimated top quark decay.
% With these issues in mind, one can try to find a jet representation which adresses both issues presented above, dealing optimally with inputs of different dimensionalities, and avoiding not physically motivated pre-processing steps. 

% These problems lead to the question of whether a more natural jet representation can be built, one which avoids the loss of information and generalizability induced by particularly the pre-processing step. 
The algorithms to be described below think about the jet instead as collection of correlated objects, each one with its set of characteristics (such as position and energy), and apply some of the sequence-based ML ideas discussed above for different types of tasks. 
This idea is reminiscent of the actual way jets are built in collider experiments, using sequential clustering algorithms (for example, the widely used in LHC experiments anti-$k_t$ algorithm~\cite{CacciariAntiKt}), acting on low-level detector quantities, such as calorimeter deposits or tracks.

% The usage of the actual jet constituents is not without its challenges. 
% For example, the number of constituents in a jet is not fixed and can vary substantially from object to object, similar to the pixel occupancy in a jet image. 
% The usual ML-based discrimination models such as densely-connected NNs and BDTs are not well-equipped to deal with these cases, which can be detrimental to performance (for example, by training a classifier with mostly masked inputs due to sparseness). 
% Therefore, architectures which are flexible enough to take data sparsity into account, and exploit it, should be investigated. 
% Among ML-based classes of models, one that satisfies this requirement is commonly known as \textit{sequence-based models}.

\subsection{Identifying Heavy Flavor Jets}
\label{sec:sequence_heavyflavor}

The identification of jets originated from the products of the hadronization process of \textit{heavy} quarks (bottom and charm quarks), known as heavy flavor jets, is of fundamental importance for experiments at the LHC for two main reasons. 
Firstly, the Higgs boson - discovered in 2012 by ATLAS and CMS, and currently the focus of intense experimental investigation - mainly decays to a pair of bottom quarks, with a predicted decay branching fraction of about $56\%$~\cite{Heinemeyer:2013tqa}. 
Secondly, the top quark, which together with the Higgs boson can help us learn about the structure of the electroweak vacuum~\cite{Buttazzo_2013}, decays almost entirely to a bottom quark plus a W boson~\cite{Zyla:2020zbs}. 

Finding these Higgs and top decays is not an easy task due to the enormous \textit{multijet} background events at the LHC (events with at least two reconstructed jets). 
These multijet events are produced with cross sections over 3 times larger than events with top quarks, and 4 times larger than events with a Higgs boson~\cite{ATL-PHYS-PUB-2019-010}. 
Fortunately, most of these events contain jets from \textit{light} quarks (quarks other than charm, botton and top) and gluons (collectively grouped as light jets). 
Therefore, learning how to separate heavy flavor signal jets from the light jets background is necessary.

\begin{figure}[ht]
    \centering
    \includegraphics[width=0.7\textwidth]{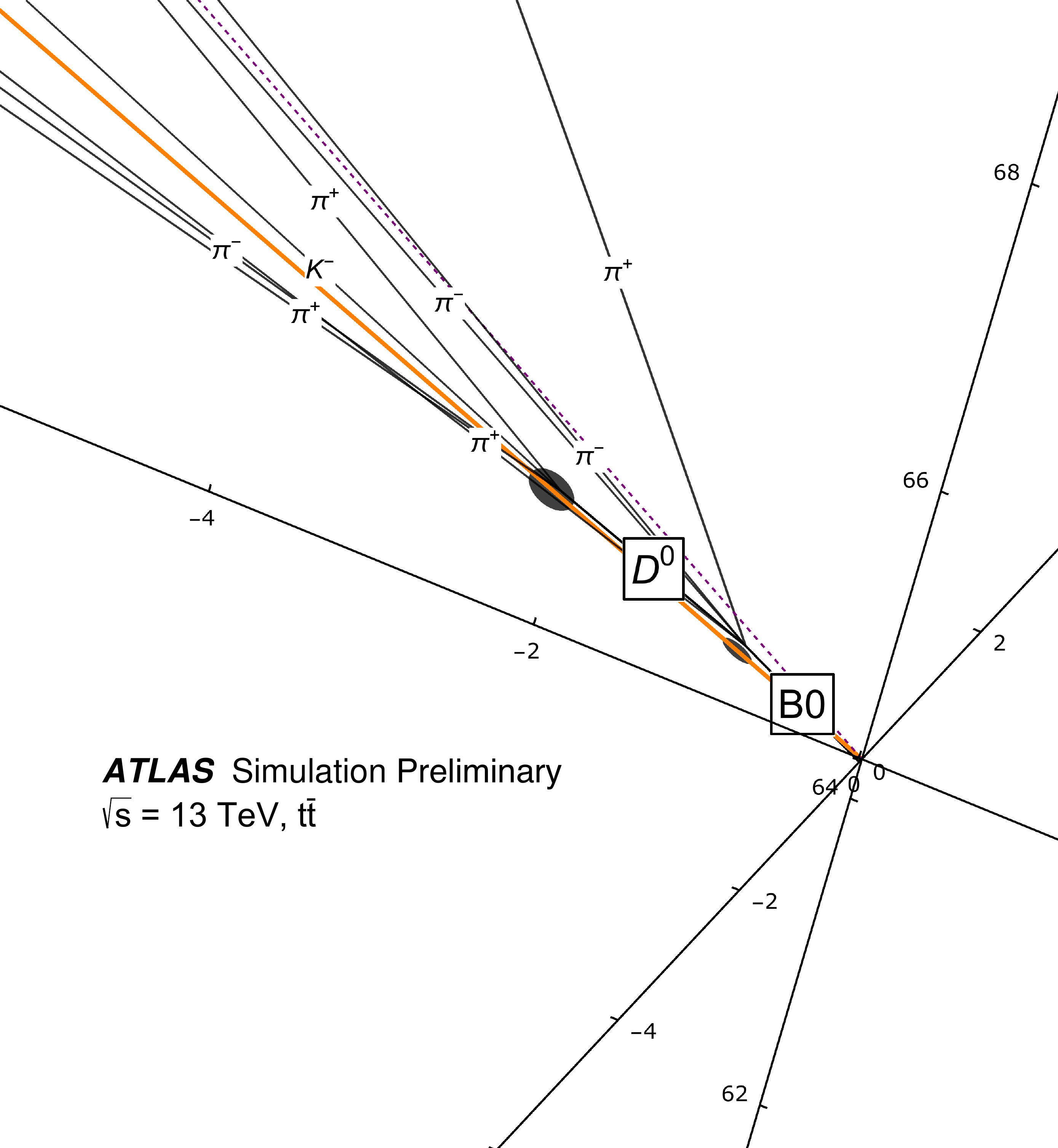}
    \caption{Simulated decay chain of a $B^{0}$ meson in the ATLAS experiment. This event display was obtained from a simulated dataset of top quark pair production, at center-of-mass energy of 13 TeV. It shows the displacement of the vertices produced from $B^{0}$ (secondary vertex) and subsequent $D^{0}$ (tertiary vertex) decays, with respect to the center of the coordinate system (primary vertex). }
    \label{fig:bhadron_decay}
\end{figure}

Hadrons containing bottom and charm quarks are heavy - for example, the $B^{0}$ meson invariant mass is about 5.3 GeV/c$^2$~\cite{Zyla:2020zbs}. 
They decay through the Weak force, thus having a long mean lifetime ($\tau_{B^{0}} \simeq 1.5\,\times\,10^{-12}$ s). 
Therefore, a $B^{0}$ with a momentum of 50 GeV/c will travel on average over 4.5 mm before decaying. 
This displaced decay can be reconstructed as a \textit{secondary vertex}, i.e., a vertex that is separated from the collision's primary vertex (where the initial proton-proton interaction ocurred, in the case of the LHC). 
The LHC experiment's inner trackers (reponsible for reconstructing the trajectory of charged particles) have been built with the intent of separating these secondary vertices with good precision, in order to identify heavy flavor jets. 
The decay chain of a $B^{0}$ meson, as simulated by the ATLAS experiment, is shown in Figure \ref{fig:bhadron_decay}~\cite{ATL-PHYS-PUB-2018-025}, which also includes the \textit{tertiary vertex} from the displaced decay of the $D^{0}$ meson, containing a charm quark. 

In general, requiring that a secondary vertex needs to be reconstructed in order to identify a b-jet can be detrimental for a high efficiency algorithm. 
However, tracks originated from this displaced location will have very particular characteristics when compared to tracks from the primary vertex, and will be correlated by their shared origin. 
Therefore, algorithms that focus on finding correlations between tracks perform comparatively well with respect to direct secondary vertex finding, and can provide complimentary information for a combined heavy versus light jet discrimination. 

In the ATLAS experiment, two different types of algorithms have been developed to identify heavy flavor jets based on the likelihood of tracks being originated from secondary vertices. 
Both use the tracks' \textit{impact parameter} information, which encodes the distance of closest approach of the charged particle's trajectory with respect to the primary vertex. 
Particles originated from the primary vertex will have small impact parameters, while particles from secondary vertices will tend to have larger impact parameters. 
An important related quantity is the impact parameter significance, in which the distance is divided by the uncertainty in its measurement. 
Utilizing the significance minimizes the impact of low-quality tracks with large mis-measured impact parameters. 

\begin{figure}[ht]
    \centering
    \includegraphics[width=0.75\textwidth]{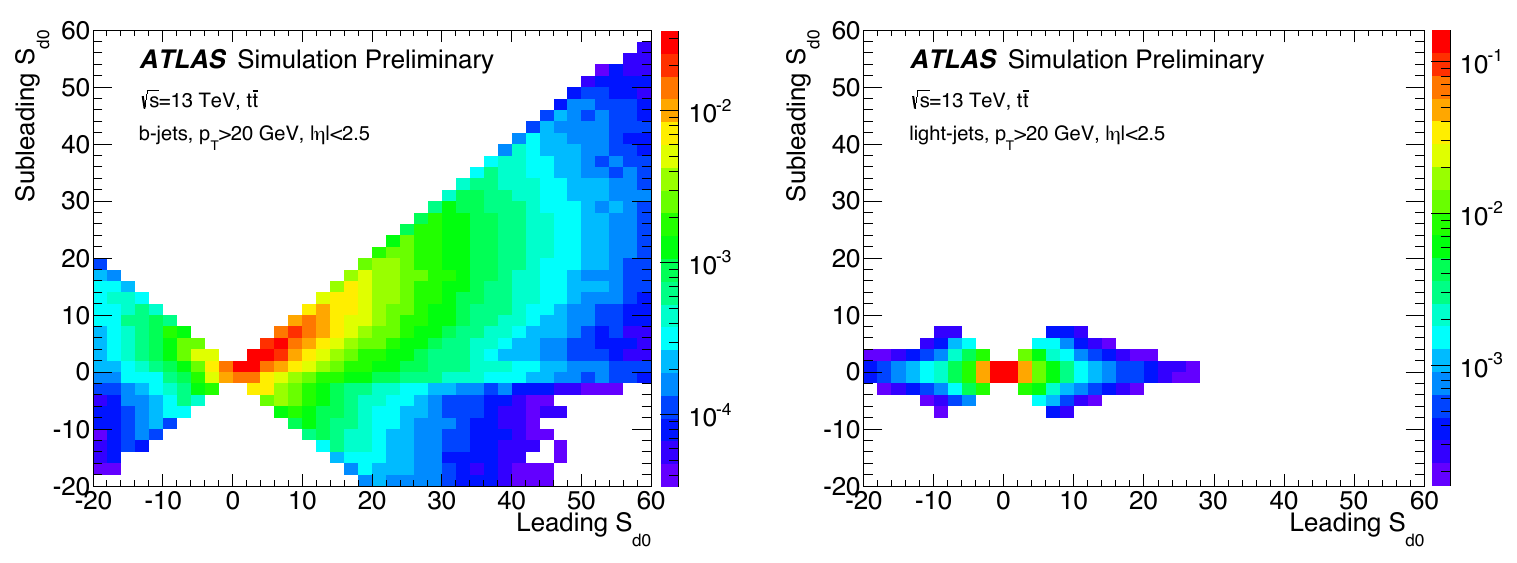}
    \caption{2D histogram of $\mathcal{S}_{d_0}$ for leading (horizontal axis) and subleading (vertical axis) tracks inside a b-jet (left) and a light flavor jet (right). 
    The correlation observed is an indication that the naïve Bayes approach of the IP3D algorithm is not enough to exploit the full information contained in the tracks' impact parameters with respect to b-jet identification. 
    }
    \label{fig:rnnip_sd0_corr}
\end{figure}

IP3D~\cite{Aad_2019_IP3D}, one of the first ATLAS algorithm based on tracks impact parameter information, treats the tracks as independent entities, ignoring possible correlations. 
It uses 3D histograms of transverse and longitudinal impact parameter significances ($\mathcal{S}_{d_0}$ and $\mathcal{S}_{z_0}$, respectively), and a track quality grade, built from simulation and separately for b-jets, c-jets and light flavor jets. 
Per-flavor conditional likelihoods are calculated from these histograms for each track in a jet.
With a naïve Bayes approach, a final jet-level likelihood is built by multiplying the individual tracks likelihoods. 
However, important information is lost with the assumption that the tracks in the jet are uncorrelated. 
This can be seen in Figure~\ref{fig:rnnip_sd0_corr}, where a strong correlation between the $\mathcal{S}_{d_0}$ distribution of the leading and subleading track in the jet (ordered by $\mathcal{S}_{d_0}$) can be seen near the diagonal for b-jets but not for light jets.

More recently, algorithms exploiting RNN architectures based on LSTMs have been proposed to perform the task described above, treating the list of tracks within the jet as the input sequence to the algorithm. 
Due to the variable number of tracks within a jet, RNNs are better suited than dense architectures in this approach. 
Even though no natural track ordering is clear to the problem, tracks with larger impact parameters are more likely to come from heavy flavor jets, therefore, impact parameter based ordering is a good ansatz\footnote{Empirically, the studies mentioned below have shown that certain track orderings work better than others.}.
While the most basic version of this algorithm acts on tracks only, proposals have been made to combine track and secondary vertices information in a single LSTM-based architecture~\cite{GuestFlavorDNN}.

The ATLAS implementation of the LSTM-based architecture for heavy jets identification with tracks' impact parameters is called RNNIP~\cite{ATLAS-RNNIP}. 
It treats the tracks within a jet as a sequence, and uses the impact parameter significance information and track kinematics as features. 
It also uses categorical information based on the track reconstruction quality in an embedded layer. 
These categories separate high quality, well measured tracks, in which a better impact parameter resolution is expected, based on detector-level information such as the number of hits in the innermost tracker layer. 

The tracks are ordered by $\mathcal{S}_{d_0}$, although other orderings (such as by track $p_T$) have shown similar performance. 
The algorithm is trained in a simulated sample of top pairs, which provide a dataset enriched of both heavy and light quarks, and outputs a probability of a given jet to be a bottom jet ($b$-jet), a charm jet ($c$-jet), or a light flavor jet. 
The three probabilities are then combined into a likelihood that is used for discrimination.  

\begin{figure}[ht]
    \centering
    \includegraphics[width=0.75\textwidth]{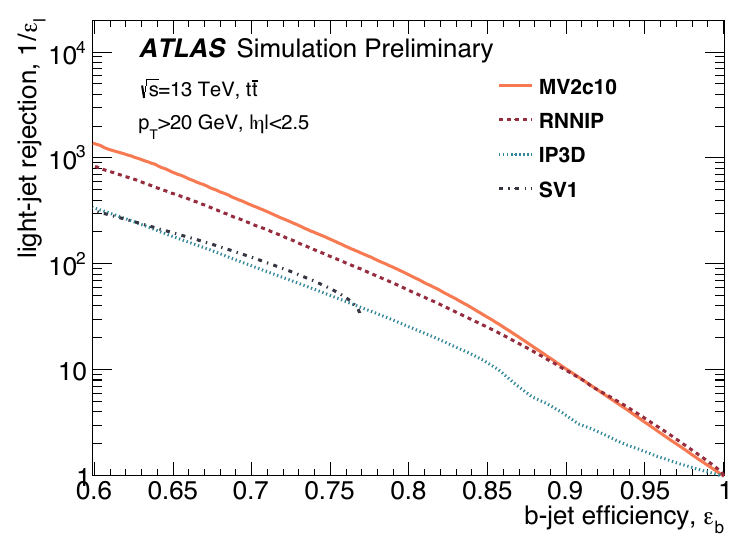}
    \caption{Performance of heavy flavor identification algorithms in the ATLAS experiment~\cite{ATLAS-RNNIP}. The horizontal axis shows the efficiency of correctly identifying $b$-jets, while the vertical axis shows one over the efficiency of incorrectly identifying light flavor jets as $b$-jets. The red dashed curve shows the performance of the RNN-based algorithm (RNNIP), while the dashed blue curve shows the performance of a naïve Bayes algorithm acting on similar inputs (IP3D). 
    }
    \label{fig:rnnip_ip3d_roc}
\end{figure}

A comparison between the performance of the naïve Bayes algorithm described above (IP3D) and RNNIP is shown in Figure \ref{fig:rnnip_ip3d_roc}. 
The efficiency of identifying $b$-jets is plotted on the horizontal axis, while one over the probability of identifying light flavor jets (misidentification probability) as $b$-jets is plotted on the vertical axis. 
The RNNIP algorithm displays a better light flavor jet discrimination for every value of $b$-jet efficiency, and is comparable to a boosted decision tree that combines IP3D with the secondary vertex-based ATLAS algorithms (MV2c10~\cite{Aad_2019_IP3D}), even though it does not explicitly reconstruct secondary vertices. 
Performances were measured for jets clustered with the anti-$k_T$ algorithm~\cite{CacciariAntiKt} with R = 0.4, with a transverse momentum above 20 GeV, in a simulated dataset of top quark pairs, at center-of-mass energy of 13 TeV.

\begin{figure}[ht]
    \centering
    \includegraphics[width=0.9\textwidth]{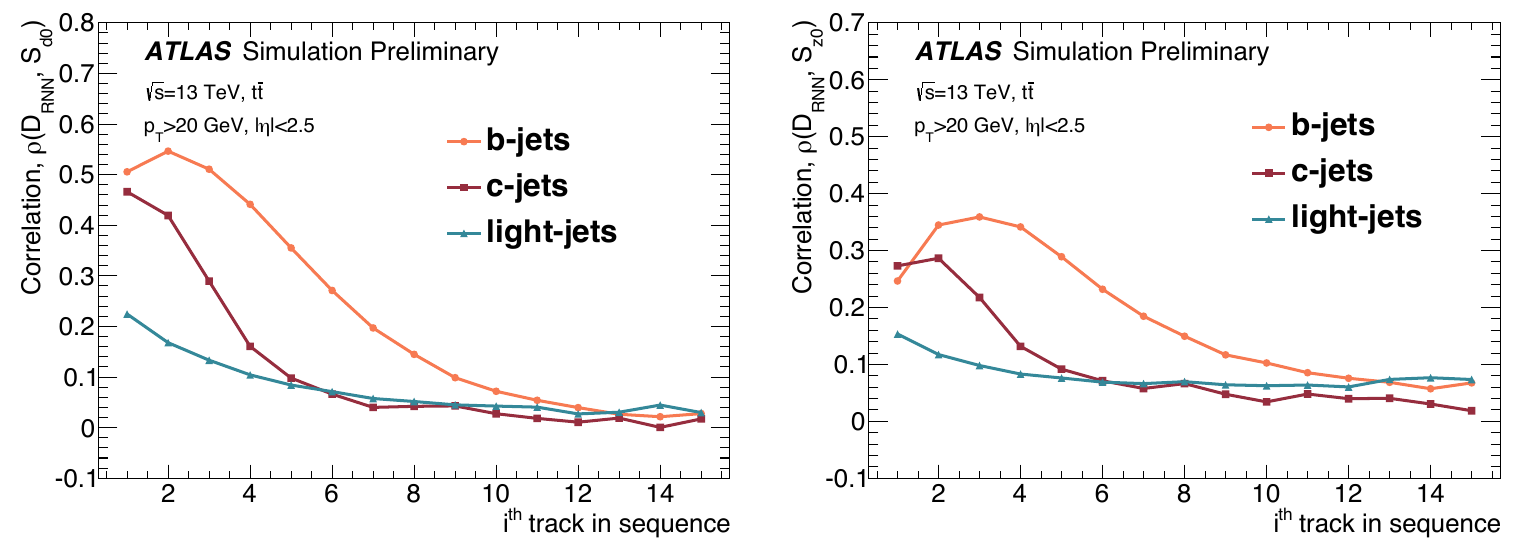}
    \caption{Pearson's correlation factor between RNNIP likelihood, and transverse and longitudinal impact parameter significances. Correlations are shown separately for b-jets (orange), c-jets (red), light flavor jets (blue).}
    \label{fig:rnnip_corr}
\end{figure}

Figure~\ref{fig:rnnip_corr} shows the Pearson's correlation coefficient $\rho$ between the RNNIP likelihood, and $\mathcal{S}_{d_0}$ and $\mathcal{S}_{z_0}$ for each track in the sequence. 
It is interesting to note that stronger correlations in b-jets are seen for  impact parameter significances of the first $\sim 8$ tracks, which may be related to the expected charged particle multiplicity of b-hadron decays. 
This shows that the network architecture is able to learn contextual information from the given sequence ordering.

Heavy flavor jets identification in the CMS experiment shares many similarities with the strategies employed by ATLAS. 
In particular, their final discriminant is also a combination of information pertaining to secondary vertexing and the set of tracks inside the jet. 
Two sets of algorithms have been developed with this intent: one set combining engineered features extracted from the jet, and one that directly uses reconstructed objects information into a neural network architecture which includes LSTM layers. 
The CMS DeepCSV algorithm~\cite{Sirunyan_2018} exemplifies the first strategy, similarly to the ATLAS MV2c10 boosted decision tree. 
It is based on a densely-connected neural network with 4 hidden layers, with inputs that are defined by other algorithms which act directly on tracks' impact parameter information and reconstructed secondary vertices.

Significant improvement has been observed by CMS by moving to the DeepFlavour algorithm~\cite{CMS-DP-2018-033}, which acts directly on these low-level observables.
The DeepFlavour network receives three sequences as inputs: a sequence of charged particles (reconstructed from tracks and calorimeter clusters), a sequence of neutral particles (calorimeter clusters with no associated tracks), and a sequence of reconstructed secondary vertices. 
Each sequence is processed by a 1D convolutional layer, which learns a shared representation that is specific for each type of sequence. 
The convolutional layer outputs are then fed to three different LSTM layers, which summarized the sequences information into three fixed length feature vector. 
These features are combined with additional jet-level information in a densely connected network, which outputs the jet flavor probabilities.

\begin{figure}[ht]
    \centering
    \includegraphics[width=0.75\textwidth]{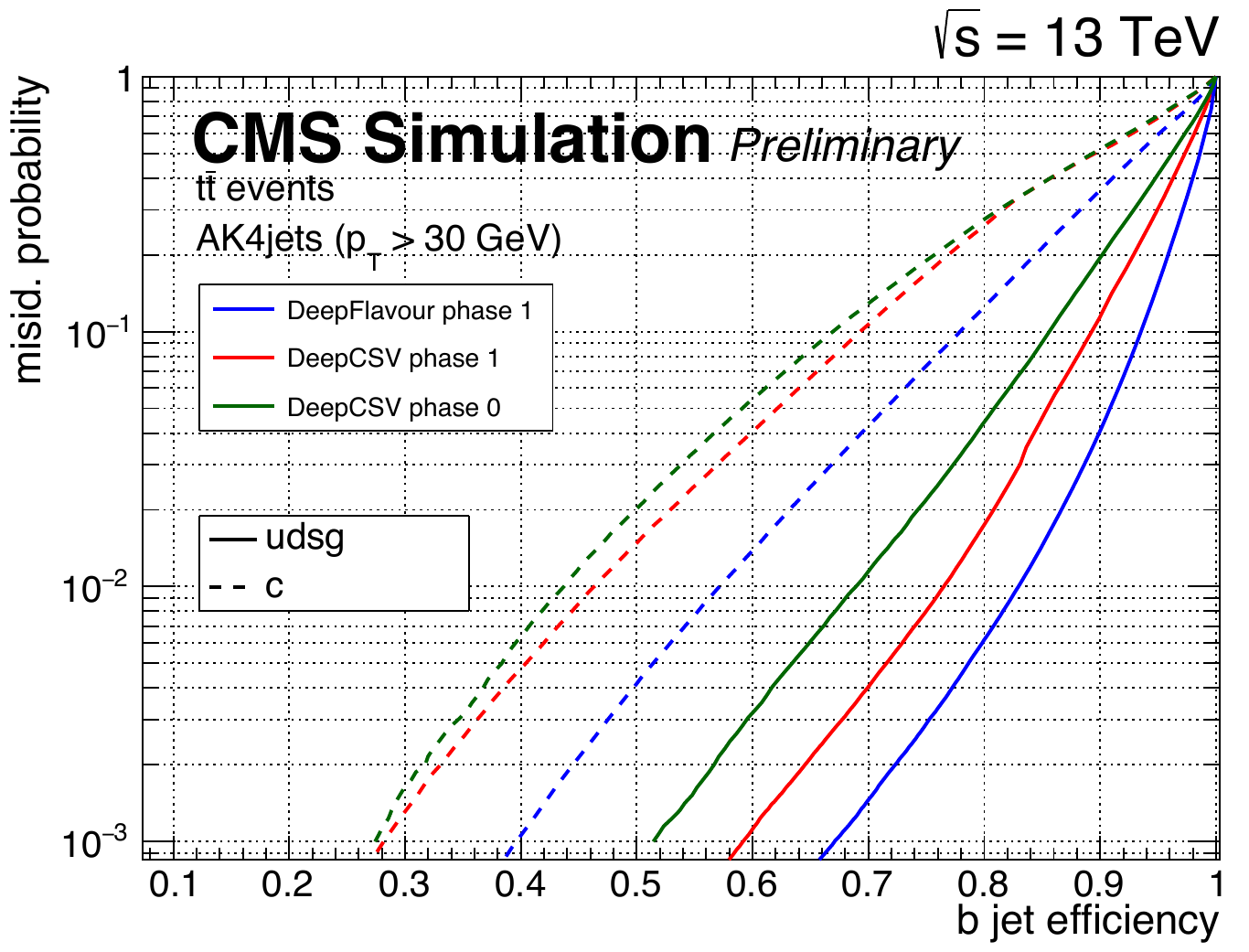}
    \caption{Performance of heavy flavor identification algorithms in the CMS experiment~\cite{CMS-DP-2018-033}. The horizontal axis shows the efficiency of correctly identifying $b$-jets, while the vertical axis shows the efficiency of identifying light flavor jets (full lines) or $c$-jets (dashed lines) as $b$-jets. The red and blue curves represent the DeepCSV and DeepFlavor algorithms performances with the 2017 CMS detector conditions, while the green curve represents the DeepCSV algorithm with the 2016 CMS detector conditions. 
    }
    \label{fig:cms_deepflavour}
\end{figure}

Figure \ref{fig:cms_deepflavour} summarizes the CMS $b$-jet identification performance. 
The red and blue lines represent the performance of the DeepCSV and DeepFlavour algorithms respectively, with the CMS detector conditions present during the 2017 LHC data taking period (Phase 1), while the green line shows the DeepCSV performance with the CMS detector conditions in 2016 (Phase 0). 
Between 2016 and 2017, CMS inner tracking detector was upgraded to deal with the harsher radiation conditions at the LHC later Run 2 years. 
This upgrade also provided the CMS experiment a better impact parameter resolution, directly improving their heavy flavor identification performance. 
The $b$-jet efficiency is shown as a function of the light flavor jets misidentification probability (full lines), and as a function of the $c$-jets misidentification probability (dashed lines). 
While a large improvement is seen in the performance of the DeepCSV algorithm with the CMS Phase 1 inner tracker upgrade, an improvement just as large is seen with the usage of DeepFlavour in terms of light flavor jet discrimination, with an even large gain in terms of $c$-jet rejection. 
Performances were measured for jets clustered with the anti-$k_T$ algorithm, with a transverse momentum above 30 GeV, in a simulated dataset of top quark pairs, at center-of-mass energy of 13 TeV.

\subsection{Identifying Strange Jets}

Jets from strange quarks are grouped within the light flavor jets category for the algorithms described above. 
However, for certain physics applications, such as the direct measurement of the $|V_{ts}|$ element of the CKM matrix through the search of the rare decay $t\rightarrow W^{+}s$ (or $\bar{t}\rightarrow W^{-}\bar{s}$)~\cite{Ali_2010_vts}, discriminating strange jets from first-generation jets is a necessity. 

A LSTM-based algorithm has recently been proposed to tackle this problem~\cite{ErdmannStrangeTagging}. 
The study is performed with a simplified detector description model (Delphes~\cite{delphes}) based on a CMS-like detector. 
The algorithm is trained to discriminate between strange jets and jets from the hadronization of up and down quarks, produced by proton-proton QCD interactions. 
Similarly to the ATLAS RNNIP, the proposed algorithm acts on sequences of tracks, with features based on their impact parameters and kinematics with respect to the jet. 

Secondary vertices are expected to be present in strange jets through the decay of strange kaons into $\pi\pi$ and lambda baryons into $p\pi$. 
Therefore, all possible secondary vertices in the jet are reconstructed by pairing tracks with small distances of closest approach to each other.
These vertices are then used to define a track ordering based on the parameter $R$ assigned to each track. 
This parameter is defined either by the transverse distance between the primary vertex and the secondary vertex to which that track belongs, or, in the cases where the track is not associated to a secondary vertex, the innermost tracker hit belonging to that track. 
If multiple tracks receive the same $R$ (e.g., two tracks from the same secondary vertex), their ordering is performed by $p_T$. 
This ordering ensures that adjacent tracks belong to the same secondary vertex, which the network will use to learn about displaced decays. 

\begin{figure}[ht]
    \centering
    \includegraphics[width=0.75\textwidth]{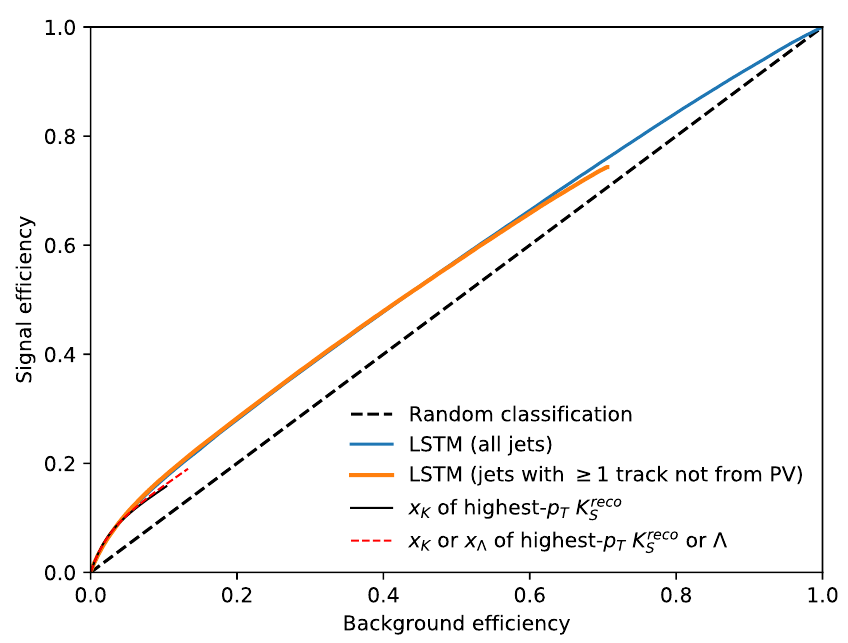}
    \caption{Performance of strange jet identification algorithms based on LSTM architectures and strange hadron reconstruction~\cite{ErdmannStrangeTagging}. 
    The vertical axis shows the efficiency of correctly identifying strange jets, while the horizontal axis shows the efficiency of incorrectly identifying up and down quark jets as strange jets. 
    The blue and orange lines show the performance of the LSTM-based algorithm using all jets and jets with at least one track with transverse impact parameter $|d_0| > 1$ mm, respectively. 
    The red dashed and full black lines show the performance of selecting on the transverse momentum fractions $x_K$ and $x_\Lambda$, and only on $x_K$, respectively.
    }
    \label{fig:strange}
\end{figure}

Figure~\ref{fig:strange} compares the performance of the LSTM-based strange jet discriminator to the performance of simpler methods. 
In particular, the performance of using the transverse momentum fraction $x_K$ and $x_\Lambda$ of identified kaons and lambda baryons is also investigated. 
To calculate these quantities, a selection on the invariant mass of the reconstructed secondary vertices, consistent with the $K$ and $\Lambda$ masses, is applied. 
The highest $p_T$ $K$ and $\Lambda$ candidates of the remaining vertices are chosen, and used to calculate $x_K$ and $x_\Lambda$ to their parent jet. 
Two different setups of the LSTM are compared: one including all selected jets, and one only including jets that contain at least one track with large transverse impact parameter $|d_0| > 1$ mm.

The overall performance of this algorithm is unfortunately limited by the similarity between strange and first-generation quark jets - achieving a background efficiency of $21\%$ ($63\%$) for a signal efficiency of $30\%$ ($70\%$).  
The secondary vertices from kaon decays are not as displaced as vertices from $b$-jets or $c$-jets and can easily be misidentified from vertices produced by material interaction or decays of particles originated in the hadronization process. 
One important improvement with respect to using $x_K$ and $x_\Lambda$, however, is the ability to achieve higher signal efficiencies, as shown in Figure~\ref{fig:strange}.

\subsection{Identifying tau lepton jets}

Similarly to heavy flavor jet identification, identification of tau leptons is particularly interesting due to its ties to Higgs physics. 
The coupling between the Higgs and the tau lepton is the largest Higgs couplings to leptons in the Standard Model. It is therefore an opportunity to directly measure the structure of the Higgs Yukawa couplings to that sector of the Standard Model.

Taus decay either leptonically ($\tau\rightarrow \nu_{\tau}+\ell\nu_{\ell}$, in which $\ell$ is an electron or a muon), or hadronically ($\tau\rightarrow \nu_{\tau}+$ hadrons). 
Leptonic tau decays are roughly indistinguishable from isolated leptons in hadron collider experiments. Therefore, tau identification focuses on hadronic taus, which represent a branching fraction of approximately $65\%$~\cite{Zyla:2020zbs}. 
Hadronic tau decays usually include one or three charged pions and one or more neutral pions. Therefore, these decays are seen in the detector as narrow jets with one or more tracks.

Since neural pions do not leave signals on the detectors' inner tracks, their trajectories cannot be reconstructed as tracks. 
This means that if only track information were used, a large portion of information for tau identification would be missing. 
Therefore, an optimal strategy should aim to combine the tracking and calorimetry information.

The ATLAS experiment state-of-the-art tau identification algorithm is based on a double LSTM architecture that combines track sequences and calorimeter deposits (clusters) sequences~\cite{ATLAS-RNNTau}. 
The algorithm has three sets of inputs: a track sequence, a cluster sequence, and a set of high-level variables connected to a dense layer. 
The track and cluster features considered refer to their kinematics and detector-level properties, while the the high-level ones are related to the jet itself, or engineered features based on the collection of jet constituents.

Tracks are individually fed through dense layers with shared weights, so that an embedded representation can be learned. 
The same procedure is applied to calorimeter clusters separately. 
The two processed sequences, one of track embeddings and one of cluster embeddings, are ordered by decreasing $p_T$ of the original objects and used as inputs to two separate LSTM blocks. 
These blocks includes two layers of LSTM units - the first one maps the sequence in the learned representation into another sequence of the same length. The second LSTM layer only outputs the information at the last time-step, thus providing a summary of the input sequence. 
The LSTM blocks outputs are fed to a densely-connected block, which also receives information from a densely-connected block encoding high-level observables of the tau jet.

The training and evaluation of the architecture defined above is performed separately for the cases in which the tau decay includes one or three tracks.  
The tau decays (signal) are provided by simulation of $\gamma^{*}\rightarrow \tau\tau$ events, while background jets are selected from a simulation of dijet events.

\begin{figure}[ht]
    \centering
    \includegraphics[width=0.75\textwidth]{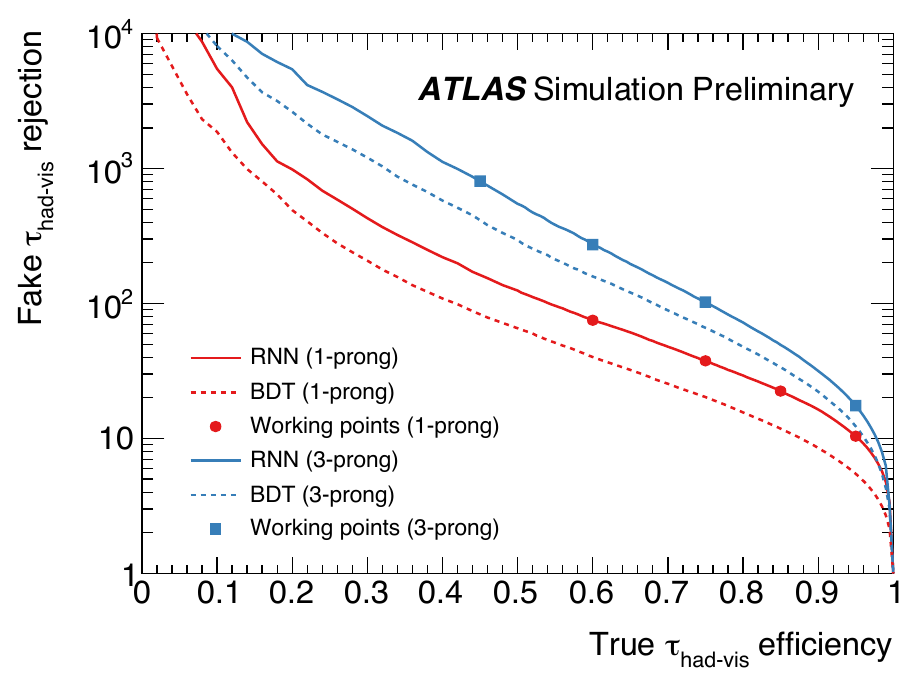}
    \caption{Performance of hadronic tau jets algorithms in the ATLAS experiment~\cite{ATLAS-RNNTau}. The horizontal axis shows the efficiency of correctly identifying hadronic tau jets, while the vertical axis shows the inverse of the efficiency of incorrectly identifying quark-initiated jets as hadronic tau jets. 
    The red curves show the performance exclusively on tau jets with a single track (1-prong), while the blue curves represent tau jets with three tracks (3-prong). 
    The RNN-based model's performance is shown in the full lines, with the dashed lines representing an algorithm with similar inputs but based on a boosted decision tree.}
    \label{fig:tauid_roc}
\end{figure}

The performance obtained in Figure \ref{fig:tauid_roc} compares the LSTM architecture (RNN) optimized for tau decays with one (1-prong) and three (3-prong) tracks. It also compares to the previous algorithm used in the ATLAS experiment, based on a boosted decision tree (dashed lines). 
The LSTM-based architecture outperforms the previous baseline for all hadronic tau efficiencies. 
These improvements have been shown to be significant enough that the new architecture has actually been used for identifying tau candidates at the ATLAS High-Level Trigger (HLT) in 2018.

\subsection{Identifying Top Jets}

When produced at large momenta, top quarks' decay products will start to merge, making it more difficult to resolve them spatially in the detector. In this regime, the entire top quark decay can be clustered into a single jet. Usually these jets have larger radius parameters: CMS~\cite{cms_largeR} utilizes R = 0.8 anti-$k_t$ jets and R = 1.5 Cambridge-Aachen jets~\cite{cajets}, while ATLAS~\cite{atlas_top} focuses on R = 1.0 anti-$k_t$ jets. 
Identifying these boosted top objects, from a background of jets from the hadronization of lighter quarks and gluons, is particularly interesting when searching for Beyond the Standard Model physics which predict TeV-range resonances decaying to top pairs. 

Several interesting features which are present in a boosted top jet can be used to discriminate against a high momentum jet produced by QCD interactions. 
In particular, hadronic top decays will tend to be three-pronged, with each prong corresponding to a final state particle in the $t\rightarrow bW\rightarrow bqq'$ decay chain. 
Two important details on this chain is that one of these prongs will be consistent with a heavy flavor jet, and the other two will be consistent with a W boson decay. 

LSTM-based architectures have been proposed for identifying these boosted top jets~\cite{EganTopTagging}. 
The study is performed with a Delphes-based detector simulation~\cite{delphes} with a particle flow type of particle reconstruction, overlaying minimum bias events to emulate the LHC 2016 collisions conditions, averaging of 23 proton-proton interaction per event. 
Jets are clustered following the ATLAS strategy, with the anti-$k_T$ algorithm and R = 1.0. 
The signal top jets are obtained from simulating a beyond the Standard Model process in which a $Z'$ boson with masses ranging from 1400-6360 GeV decays to hadronically decaying $t\bar{t}$ pairs. 
Background jets come from the simulation pure QCD hard scattering processes (QCD jets).

\begin{figure}[ht]
    \centering
    \includegraphics[width=0.75\textwidth]{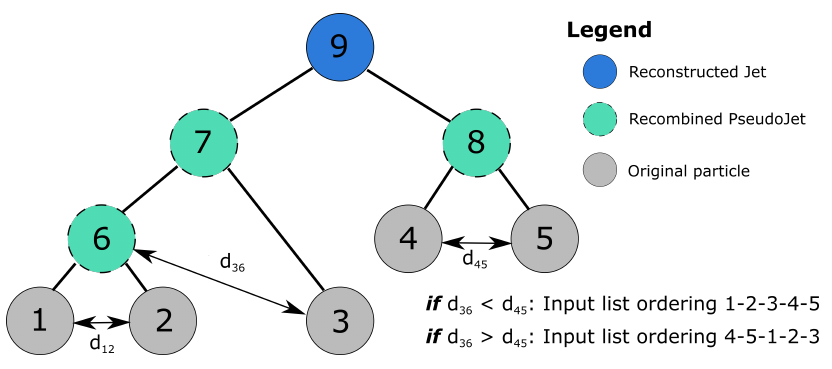}
    \caption{
        Scheme of jet clustering history used to define input sequence ordering in LSTM-based top identification algorithm~\cite{EganTopTagging}. 
        The anti-$k_T$ distance $d_{ij}$ is used to choose which path to follow in the tree. 
    }
    \label{fig:top_ordering}
\end{figure}

Similar to tau identification, the sequence for the recurrent model is built of calorimeter clusters. 
The input ordering is defined based on going through the clustering history of the jet, starting from the final reconstructed jet, and adding constituents to the sequence as they appear in each step. 
The decision on which path to follow in each time two parent nodes merge depends on the anti-$k_T$ distance metric $d_{ij}$. 
If the two parent nodes are present in the list of jet constituents, they are added to the sequence ordered by $p_T$.
A scheme presenting how the clustering tree is used to build the sequence ordering is shown in Figure~\ref{fig:top_ordering}. 
Another strategy is based on reconstructing R = 0.2 anti-$k_T$ subjets, order them by descending $p_T$, then adding constituents to the sequence depending to which subjet they belong - constituents on the same subjet are also ordered by descending $p_T$.
These schemes are compared with purely odering on the jet constituents $p_T$. 

\begin{figure}[ht]
    \centering
    \includegraphics[width=0.75\textwidth]{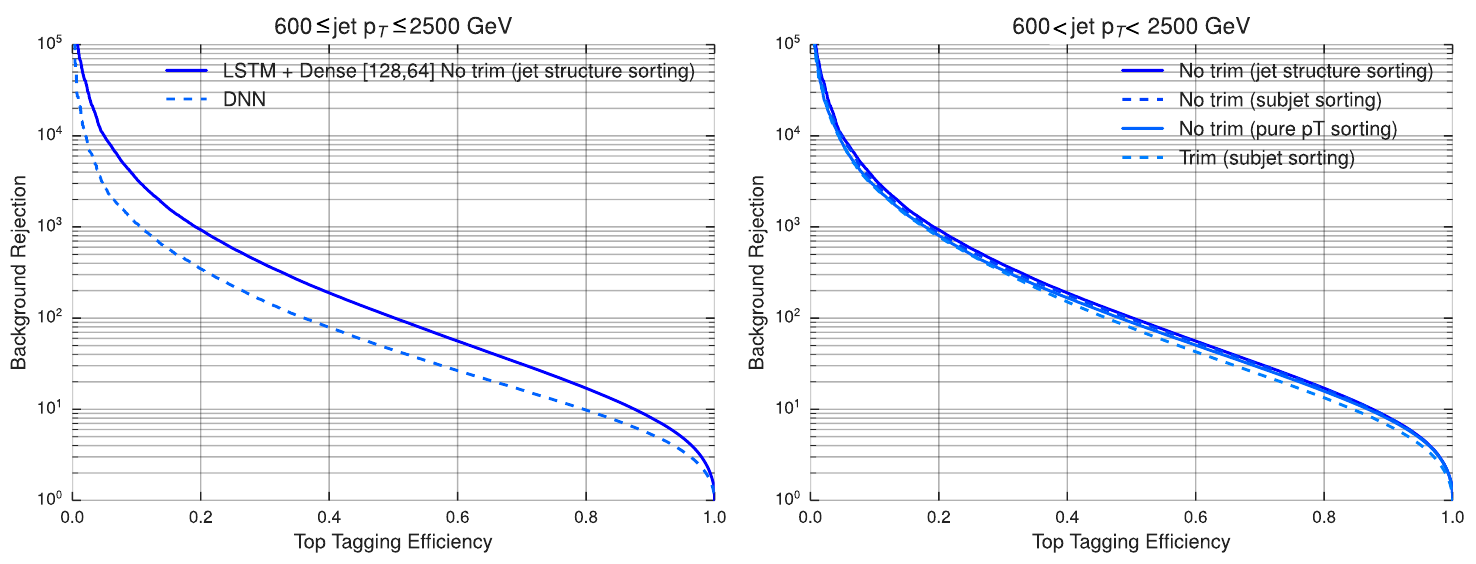}
    \caption{
        Performance of top identification with LSTM-based techniques ~\cite{EganTopTagging}. 
        The horizontal axis shows the efficiency of correctly identifying top jets, while the vertical axis shows one over the efficiency of incorrectly identifying QCD jets as top jets. 
        The left plot compares the LSTM algorithm with a strategy using a dense neural network. 
        The right plot compares different training strategies for the LSTM, including different sequence oderings (substructure, subjet or purely constituent $p_T$ based) and with or without trimming. 
    }
    \label{fig:top_tag}
\end{figure}

As seen in Figure~\ref{fig:top_tag} (left), the LSTM-based architecture with substructure oredering outperforms the previous baseline studied by the same group~\cite{top_dnn}, based on a fully connected dense neural network. 
The results are presented depending on the ordering scheme as well whether the jets are trimmed or not. 
Trimming~\cite{trimming} is a technique that ensures robustness of the jet kinematics, especially the jet mass, with respect to pile-up contamination by removing R = 0.2 subjet constituents with a certain energy fraction; in this case, subjets are removed if their energy fraction is lower than $5\%$. 
Figure~\ref{fig:top_tag} (right) shows that the impact of the ordering strategy is limited. In fact, it appears to be smaller than the impact of applying trimming, which removes subjet information that could be important for top identification. 
For more information on top jet identification with jet images, see section XYZ.

\subsection{Learning a jet representation}

In general, a jet can be understood as a collection of final state particles that are results of the hadronization and fragmentation processes of a quark or a gluon. 
However, the actual procedure of reverse-engineering these processes, which are quantum mechanical by nature, can be complicated, especially when dealing with the busy environment of a hadron collider event. 
Jet clustering algorithms aim to reduce this complexity and make the connection between observable final state particles and the phenomenological predictions based on partons. 

While many types of jet clustering algorithms have been proposed in the past, most recent applications utilize strategies based on sequential clustering, such as the ones already discussed here ($k_T$, anti-$k_T$ and Cambridge-Aachen being three well-known examples). 
These algorithms take advante from the fact that the processes for the particle shower development can be approximated well, due to factorization theorems, by a sequence of $2\rightarrow 1$ splittings in creating a hierarchical representation of the jet. 
When reverse-engineered, this sequence of splittings becomes a sequence of mergings of jet constituents, forming a binary tree. 
This binary tree represents the jet clustering history, and encodes important information about the nature of that jet. 

As presented in the previous section, the usage of RNN architectures for learning jet labels has been a successful avenue in collider experiments, greatly improving on previous, often already Machine Learning-based, strategies. 
Treating the jet constituents simply as an input sequence, however, des not fully exploit the full information contained in the jet clustering history as represented by the clustering tree. 

It has been suggested that through understanding the interplay between the clustering history and neural network architectures could lead to a more natural jet representation~\cite{AndreassenJUNIPR}. 
The study presents a framework in which a neural network can be built based on the clustering tree obtained from the jet clustering algorithm history. 
It tries to exploit the entire physical knowledge contained within the jet, particularly with respect to hierarchical correlations between individual constituents. 
This structure is then used to learn an overall probabilistic model of the jet conditioned on its constituents, through an unsupervised training task.

Building a probabilistic model for a class of jets presents the possibility of having a tractable and differentiable distribution function which can inform different aspects of jet physics. 
In particular, models trained for different classes of jets, such as b-jets and light flavor jets, can be used to compute likelihood ratios for discrimination. 
Models can also be sampled in order to generate new jets, a process which can take a significant amount of time when large datasets are required. 

\begin{figure}[ht]
    \centering
    \includegraphics[width=0.75\textwidth]{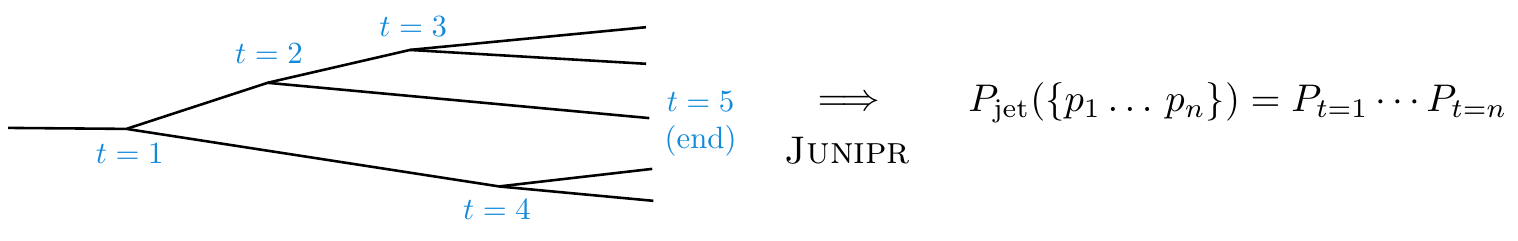}
    \caption{JUNIPR calculation of jet-level probability model based on its factorization to nodes in binary tree of the jet clustering history~\cite{AndreassenJUNIPR}. }
    \label{fig:junipr_prob}
\end{figure}

The framework, named JUNIPR ("Jets from UNsupervised Interpretable PRobabilistic models"), is based on the factorization of the jet probabilistic model into a product of probabilities given by each step of the clustering tree. 
For a set of jet constituents denoted by their 4-momenta $p_1,\ldots,p_n$, it computes the probability density $P_{\text{jet}}(\{p_1,\ldots,p_n\})$ of this set to have arisen from the specified model. 
This probability factorizes based on the $2\rightarrow 1$ clustering tree, so that $P_{\text{jet}}(\{p_1,\ldots,p_n\}) = \prod_t^{t=n} P_{t}$, where $P_t$ represents the probability model of the branching step $t$. 
This factorization is schematically shown in Figure~\ref{fig:junipr_prob}. 

The study further models $P_t$ as the product of three independent probabilities: $P_{\text{end}}$, probability over binary values predicting if the tree stops after this split; $P_{\text{mother}}$, probability over integers of how likely is that the mother in step $t$ indeed participates on the splitting $1\rightarrow 2$; and $P_{\text{branch}}$, probability over kinematic configurations of the three states involved in the branching step $t$. 
Each one of these three probabilities are modeled with densely-connected neural networks. 
They use as inputs the hidden state $h_t$ at branching step $t$, calculated based on the recurrent relation below:
\begin{equation}
h_t = \tanh{ \left( V \cdot \left( k_{d_1}^t, k_{d_2}^t\right) + W \cdot h_{t-1} + b \right) },
\label{eq:junipr}
\end{equation}
where $k_{d_1}^t, k_{d_2}$ represent the 4-momenta of the two daughters involved in the splitting step $t$, and $V$, $W$, and $b$ represent learnable weights and biases. 
This equation has the same form as the one presented in section \ref{ssec:rnn} detailing the functioning of RNNs. 
The study also tried to replace this simple RNN strategy in the equation above with more complex solutions, such as LSTMs and GRUs. 
No significant improvement in performance was observed; it was argued that this was due to the simplicity of the task, acting on sequences with only two elements. 

The algorithm is trained based on the full jet probabilistic model, maximizing the log-likelihood over the examples in the training dataset, with stochastic gradient descent. 
Datasets are generated by $e^{+}e^{-}$ collisions, without detector simulation, at center-of-mass energy of 1 TeV. 
The two jets in the final state are separated into hemispheres by the exclusive $k_T$ algorithm~\cite{exkt}, and their clustering trees are obtained with the Cambridge-Aachen algorithm. 

\begin{figure}[ht]
    \centering
    \includegraphics[width=0.75\textwidth]{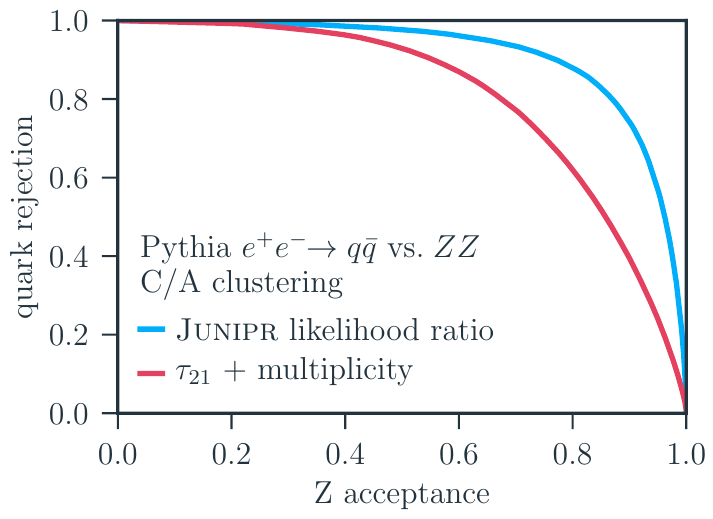}
    \caption{Performance of JUNIPR used for binary classification, based on the calculation of likelihood ratios~\cite{AndreassenJUNIPR}. 
    The horizontal axis shows the efficiency of identifying Z jets, while the vertical axis shows one minus the efficiency of incorrectly identifying quark jets as Z jets. 
    The image also shows the discrimination performance by simply using a substructure variable $\tau_{21}$ and constituents multiplicity information.}
    \label{fig:junipr_perf}
\end{figure}

Results of using the learned jet probability applied to jet classification, through likelihood ratio computation, are shown in Figure~\ref{fig:junipr_perf}. 
The blue line represents the discrimination power between jets from quarks and jets from a boosted hadronic Z decay by calculating the likelihood ratio between the jets probabilities based on each hypotheses $P_Z(\text{jet})/P_q(\text{jet})$. 
$P_Z(\text{jet})$ and $P_q(\text{jet})$ represent the JUNIPR probabilities trained individually on Z jets and quark jets, respectively, and evaluated on a given jet. 
The JUNIPR likelihood ratio greatly outperforms the strategy based on engineered features representing the jet substructure. 
 
% The scheme in Figure \ref{fig:junipr} shows how the clustering tree structure is encoded into the RNN architecture. 

% \begin{figure}[ht]
%     \centering
%     \includegraphics[width=0.75\textwidth]{figures/junipr}
%     \caption{Scheme of how the JUNIPR framework encodes a jet's clustering tree into a RNN architecture, where each hidden state learns about individual factors of a total jet probability. At each clustering step $t$, the RNN hidden state is fed the momenta of the two daughters originated from that split ($k^{(t)}_{d_1}$ and $k^{(t)}_{d_2}$). }
%     \label{fig:junipr}
% \end{figure}

The framework has also been used for direct binary classification tasks~\cite{AndreassenBinaryJUNIPR}. 
In this case, two JUNIPR networks are built based on two different types of jets (quark jets and gluon jets, for example).
The two networks are trained with a cross entropy objective function, where the individual jet probabilities from each jet type are defined by the JUNIPR networks. 
The quark versus gluon discrimination achieved by this method was seen to outperform standard approaches, such as CNNs on jet images. 
It also significantly outperforms the strategy above of individually training JUNIPR models and calculating their likelihood ratio. 

\section{Alternatives to RNNs}

The applications of RNN architectures in collider experiments and phenomenological studies has been an active area of recent developments, successfully avoiding issues previously seen with CNN architectures, while very often improving on their performances. 
Some features of these architectures are, however, less desirable for certain problems. 
Certain the physics problems utilizing RNNs do not have a well-defined, natural ordering of the sequence elements. 
Choosing a specific order becomes a non-trivial step of the data formatting, one that could lead to undesirable performance losses. 
On the other hand, other physics problems might benefit from topological structures that encode more complex relations than a sequence.

\subsection{Recursive Neural Networks}

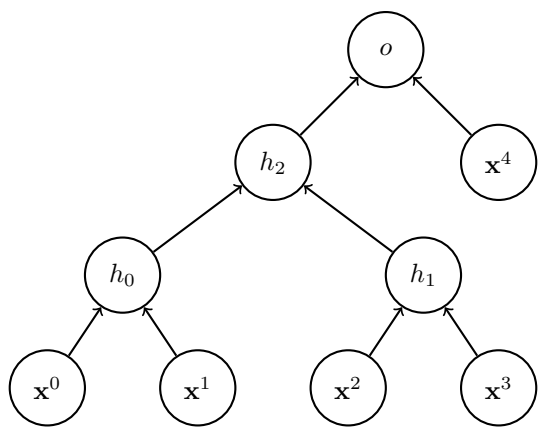
\begin{figure}
\begin{center}
\begin{tikzpicture}[
    roundnode/.style={circle, draw=black, fill=white, thick, minimum width=1.0cm},
]

\node[roundnode] (x0) at (-3,0) {$\mathbf{x}^{0}$};
\node[roundnode] (x1) at (-1,0) {$\mathbf{x}^{1}$};
\node[roundnode] (x2) at ( 1,0) {$\mathbf{x}^{2}$};
\node[roundnode] (x3) at ( 3,0) {$\mathbf{x}^{3}$};

\node[roundnode] (h0) at (-2,1.5) {$h_{0}$};
\node[roundnode] (h1) at ( 2,1.5) {$h_{1}$};

\node[roundnode] (h2) at (0,3) {$h_{2}$};
\node[roundnode] (x4) at (3,3) {$\mathbf{x}^{4}$};

\node[roundnode] (o0) at (1.5,4.5) {$o$};

\draw[->, thick] (x0) -- (h0) ;
\draw[->, thick] (x1) -- (h0) ;

\draw[->, thick] (x2) -- (h1) ;
\draw[->, thick] (x3) -- (h1) ;

\draw[->, thick] (h0) -- (h2) ;
\draw[->, thick] (h1) -- (h2) ;

\draw[->, thick] (h2) -- (o0) ;
\draw[->, thick] (x4) -- (o0) ;

\end{tikzpicture}
\caption{Scheme of a recursive binary tree structure algorithm acting on a sequence of inputs $\mathbf{x}^{0}, \mathbf{x}^{1}, \mathbf{x}^{2}, \mathbf{x}^{3}, \mathbf{x}^{4}$. The $h_i$ nodes correspond to hidden states performing the combination of two other nodes, while the $o$ node represents the output. }
\label{fig:rec-mod}
\end{center}
\end{figure}

Recursive Neural Networks (RecNNs) have been proposed in the literature as possible generalizations of RNNs, in which the sequential computational graph is replaced by a tree structure~\cite{recursive_bottou}. 
This means that a node hidden state still depends on the previous step in the computational graph, similarly to the RNN, but the step itself is defined by a binary tree instead of a sequence. 
This feature opens up the possibility of adding extra domain knowledge information into the network architecture itself when building the tree. 
A scheme of a recursive structure based on a binary tree is shown in Figure \ref{fig:rec-mod}. 

When building a RecNN, entries in a sequence are combined via a learned function, with a predetermined binary tree structure. 
Therefore, a hidden state combining entries $\mathbf{x}^{i}$ and $\mathbf{x}^{j}$ is represented by $h = f(\mathbf{x}^{i}, \mathbf{x}^{j}, \mathbf{\theta})$, where $\theta$ represents the learnable parameters. 
This function $f(.,.,\theta)$ is then used in all binary combinations, which allows the architecture to act on variable-length sequences\footnote{This is another instance of weights sharing in neural networks, a necessity when dealing with variable length inputs.}. 
Another advantage of RecNNs over RNNs is its lower complexity, since the computations are not performed per sequence entry anymore. 

Studies utilizing RecNNs with trees reproducing the jet clustering history as a basis for jet representation have been performed in the context of jet classification~\cite{LouppeQCDAwareRecursiveNN,ChengRecNNforQG,FraserJetCharge} and will be detailed below. 
They show that RecNNs are able to learn a fixed-length jet embedding from a variable length tree structure built from the jet constituents. 
This embedding can be further used for different tasks, such as classification and regression. 

%%%%%%%%%%%%%%%%%%%%%%%%

\subsubsection*{Applications to Jet Physics}

%%%%%%%%%%%%%%%%%%%%%%%% W tagging/event wide classification

\begin{figure}[ht]
    \centering
    \includegraphics[width=0.75\textwidth]{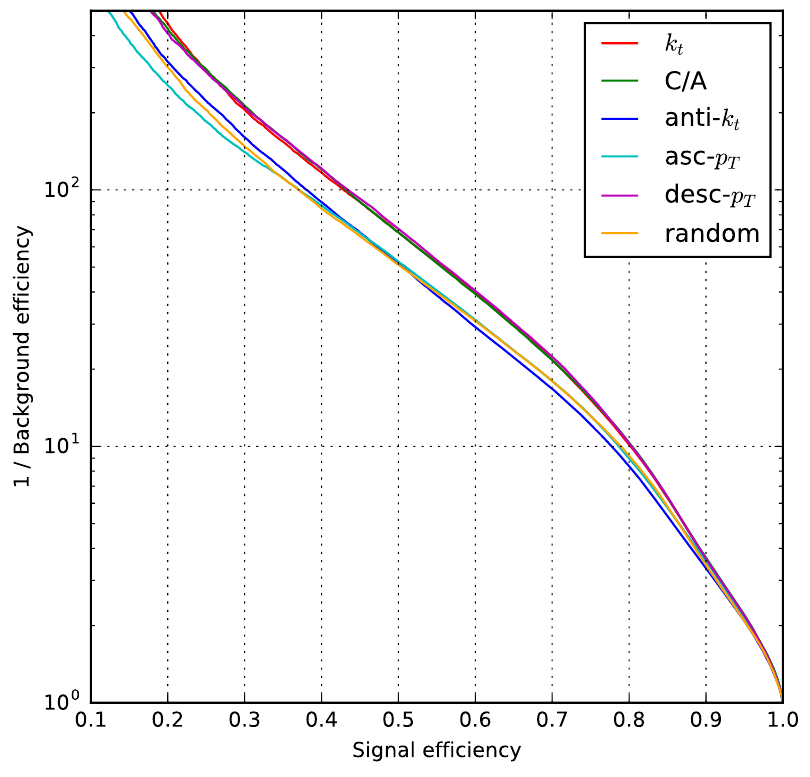}
    \caption{Performance of RecNN-based algorithm for identifying boosted W jets with respect to jets produced via purely QCD interactions~\cite{LouppeQCDAwareRecursiveNN}. 
    The horizontal axis shows the efficiency of correctly identifying W jets, while the vertical axis shows one over the efficiency of incorrectly identifying QCD jets as W jets. 
    The different line colors correspond to different jet clustering algorithms used to build the binary tree topology in the RecNN.}
    \label{fig:recnn_w}
\end{figure}

Initial studies used the RecNN architecture for jet discrimination~\cite{LouppeQCDAwareRecursiveNN}, using simulated events processed through a simplified detector simulation (Delphes). 
The signal jets are comprised of hadronically decaying W bosons reconstructed as a single R = 1.0 anti-$k_T$ jet (boosted jet), while the background is taken from purely QCD hard scattering proton-proton collisions. 
The study performs a comparison between RecNN architectures based on different binary trees, corresponding to different jet clustering algorithms histories. 
The result of this comparison is shown in Figure~\ref{fig:recnn_w}. 
It's interesting to note that, even though the jets have been initially clustered with the anti-$k_T$ algorithm, other clustering histories, such as $k_T$, perform better. 
This is consistent with previous observations that $k_T$ outperforms anti-$k_T$ in terms of identifying substructures in jets. 
In general, this variation in performance is an evidence for the strong dependence of RecNNs architectures on the choice of binary tree topology. 

The performance is also studied when adding gating to the RecNN nodes.
Similarly to LSTMs and GRUs, gated structures are used to regulate how much information is passed through the binary tree. 
An improvement in performance is observed with gating, but the importance of topology is still dominant in the results.

This study also shows that the learned RecNN jet embedding can be used for event-wide discrimination, identifying beyond the Standard Model processes involving boosted W and Z jets from purely QCD processes. 
In this case, these RecNN embeddings of the individual jets in an event are passed through as a $p_T$-ordered sequence to a GRU-based architecture which performs the classification task. 
It was observed that the usage of RecNN-based embeddings improves significantly the event classification performance with respect to using a GRU-only architecture acting on sequences of jets. 

%%%%%%%%%%%%%%%%%%%%%%%% q vs g

A similar strategy has also been employed for quark versus gluon jet discrimination~\cite{ChengRecNNforQG}, showing a slight improvement on a baseline boosted decision tree-based algorithm. 
Quark versus gluon discrimination is an important avenue of work at the LHC, due to the enormous gluon background from soft hadronic interactions, which have a strong impact on analyses with light quarks in the final state. 
One important application of this class of algorithms is identifying the hard scatter final state light jets involved in production of the Higgs boson via vector boson fusion.

\begin{figure}[ht]
    \centering
    \includegraphics[width=0.75\textwidth]{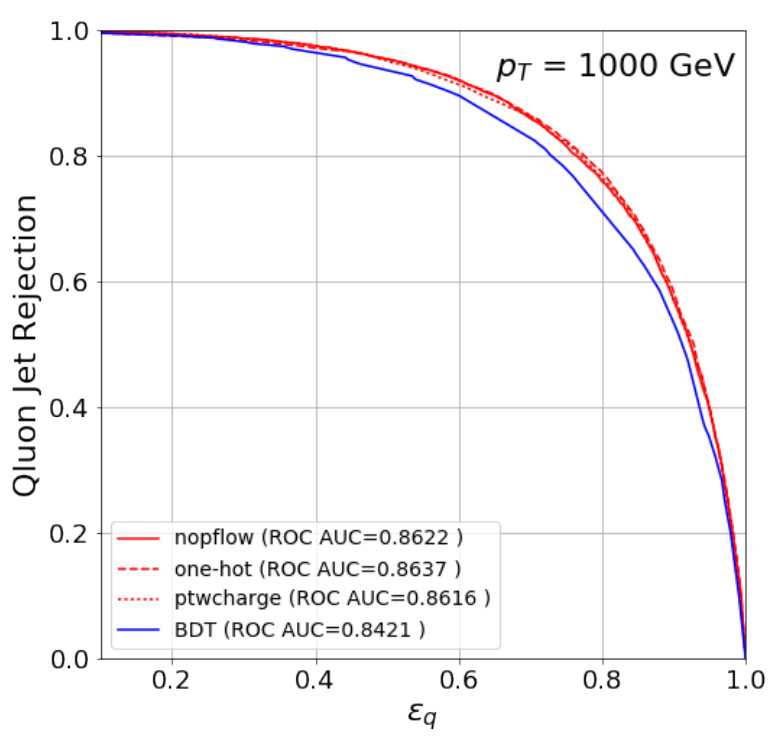}
    \caption{Performance of RecNN-based algorithm for discriminating quark and gluon jets~\cite{ChengRecNNforQG}. 
    The horizontal axis shows the efficiency of correctly identifying quark initiated jets, while the vertical axis shows one minus the efficiency of incorrectly identifying gluon jets as quark jets. 
    The red lines correspond to the RecNN models trained with different jet reconstruction techniques. The blue line shows the performance of a baseline BDT based on engineered features related to jet shape and kinematics.}
    \label{fig:recnn_qvg}
\end{figure}

As in the previous study, simulated events are processed with Delphes and jets are clustered with the anti-$k_T$ algorithm with R = 0.7 for high jets with $p_T > 1$ TeV and R = 0.4 for jets with lower $p_T$. 
Purely QCD events with two jets from the hard scatter (dijet events) are produced separately for when the hard scatter partons are gluons (background), or up, down or strange quarks (signal). 
The discrimination achieved with the RecNN under different jet reconstruction strategies is shown in Figure~\ref{fig:recnn_qvg}, together with a baseline approach based on a BDT with engineered features (jet shape and kinematics). 
Three types of jet reconstructions are compared: using calorimeter towers only ("nopflow"); using particle identification for neutral hadrons, photons, and positively and negatively charged particles, encoded in one-hot vectors for each jet constituent in the tree ("one-hot"); using a $p_T$ weighted jet charge defined by the clustering tree ("ptwcharge"). 
Although little difference is observed with respect to the investigated jet reconstruction schemes, a significant improvement is obtained over the BDT baseline.

%%%%%%%%%%%%%%%%%%%%%%%% jet charge

Recent studies have also compared RecNNs to RNNs and CNNs when tasked to estimate jet charges~\cite{FraserJetCharge}.
In the same spirit of identifying jet flavor, identifying jet charges can help with mitigating the enormous multijet background present in hadronic LHC searches. 
Requiring that two jets forming a neutral resonance have opposite charges could potentially reach that goal, assuming a good charge reconstruction resolution is achieved.

For these studies, jets from up quarks were used as proxies for positively charged jets, while jets from down quarks were used for negatively charged jets. 
Jets were simulated from QCD hard scattering processes in proton-proton interactions, and clustered from final state particles with the anti-$k_T$ algorithm with R = 0.4. 
To test the performance of CNN's on jet images, the jets were formated into $\delta\phi \times \delta\eta = 33 \times 33$ pixel box, with pixel intensities refering to the transverse momenta going into that specific pixel, and to the sum of track charges weighted by their momenta over the tracks corresponding to that pixel.

\begin{figure}[ht]
    \centering
    \includegraphics[width=0.75\textwidth]{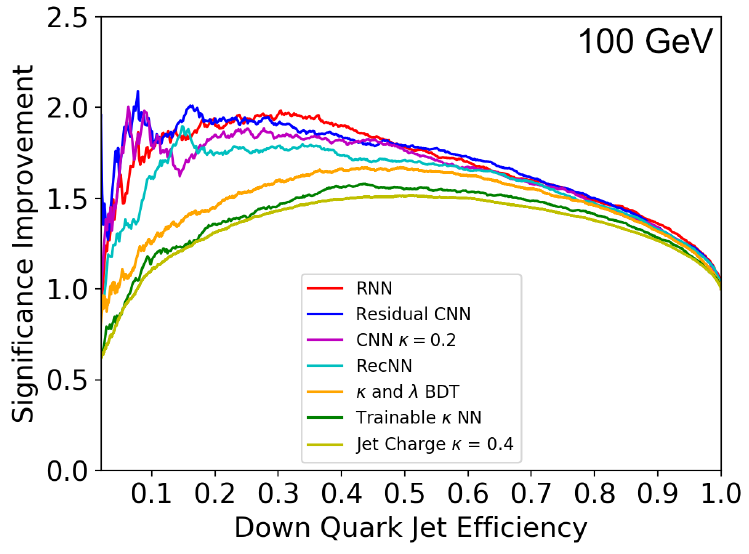}
    \caption{Performance of ML-based algorithms acting on jet constituents to discriminate up and down jets based on their electric charges~\cite{FraserJetCharge}. 
    The horizontal axis shows the efficiency of correctly identifying down quark initiated jets ($\epsilon_{s}$), while the vertical axis shows the ratio $\epsilon_{s}/\sqrt{\epsilon_{b}}$, where $\epsilon_{b}$ is the efficiency of incorrectly identifying up quark jets as down jets. 
    }
    \label{fig:recnn_charge}
\end{figure}

Results are shown comparing different ML-based models acting on jet constituents: standard CNNs, residual CNNs, RNNs and RecNNs. 
The RNN architecture implemented is based on GRUs, with LSTMs performing similarly; the jet constituents are ordered in the input sequence by their $p_T$, with the ordering based on distance to the jet axis performing equally. 
These different algorithms' performances are presented in Figure~\ref{fig:recnn_charge} in terms of Significance Improvement Curves (SIC), defined by $\epsilon_{s}/\sqrt{\epsilon_{b}}$, where $\epsilon_{s}$ is the efficiency of correctly identifying down quark initiated jets, and $\epsilon_{b}$ is the efficiency of incorrectly identifying up quark jets as down jets. 
They are compared to more standard strategies based on engineered features. 
Overall, the algorithms acting on jet constituents consistently outperform the baseline approaches. 

\subsection{Transformers and Deep Sets}

As shown by some of the studies described above, the ordering choice can be detrimental to the RNNs performance. 
One recent strategy proposed to overcome this ordering dependency is attention mechanisms~\cite{bahdanau2016neural}. 
The main working point of attention mechanisms can be understood with a simple NLP example.
Within translation problems, two languages often display different semantic structures for the same sentence: ``a yellow cat" in English becomes ``um gato amarelo" in Portuguese, with the words for ``cat" and ``yellow" switching positions. 
This relationship is difficult to be learned through standard RNNs in which the input and output sequences have a defined order. 
To solve this issue, network architectures with attention will directly learn correlations between the entries in the input sequence and the entries in the output sequence, which might not be encoded in the sequences ordering.

Different strategies and architectures involving ideas related to neural network attention have been proposed to mitigate the ordering issue. 
In particular, Transformer networks~\cite{vaswani2017attention}, which employ a self-attention technique, learning two-by-two correlations between the sequence inputs themselves through attention weights. 
Transformers have become common tools in NLP, with pretrained models such as BERT (Bidirectional Encoder Representations from Transformers)~\cite{devlin2019bert} being adopted by Google in its search engine for better understanding search queries.

Another method which aims to exploit correlations on a variable-length input structure is the Deep Sets architecture~\cite{zaheer2018deep}.
Deep Sets are particularly suited for situations in which the ordering is not well defined, as it treats the variable-length input sequence as a permutation invariant set. 
A similar Deep Sets architecture was used~\cite{Komiske_2019_PFN} to learn representations for events in collider experiments (Particle Flow Networks). 
These Particle Flow Networks act on lists of particles, such as jet constituents, learns a combined representation through dense layers with shared weights, and sums them into a single, object-wide representation. 

\begin{figure}[ht]
    \centering
    \includegraphics[width=0.75\textwidth]{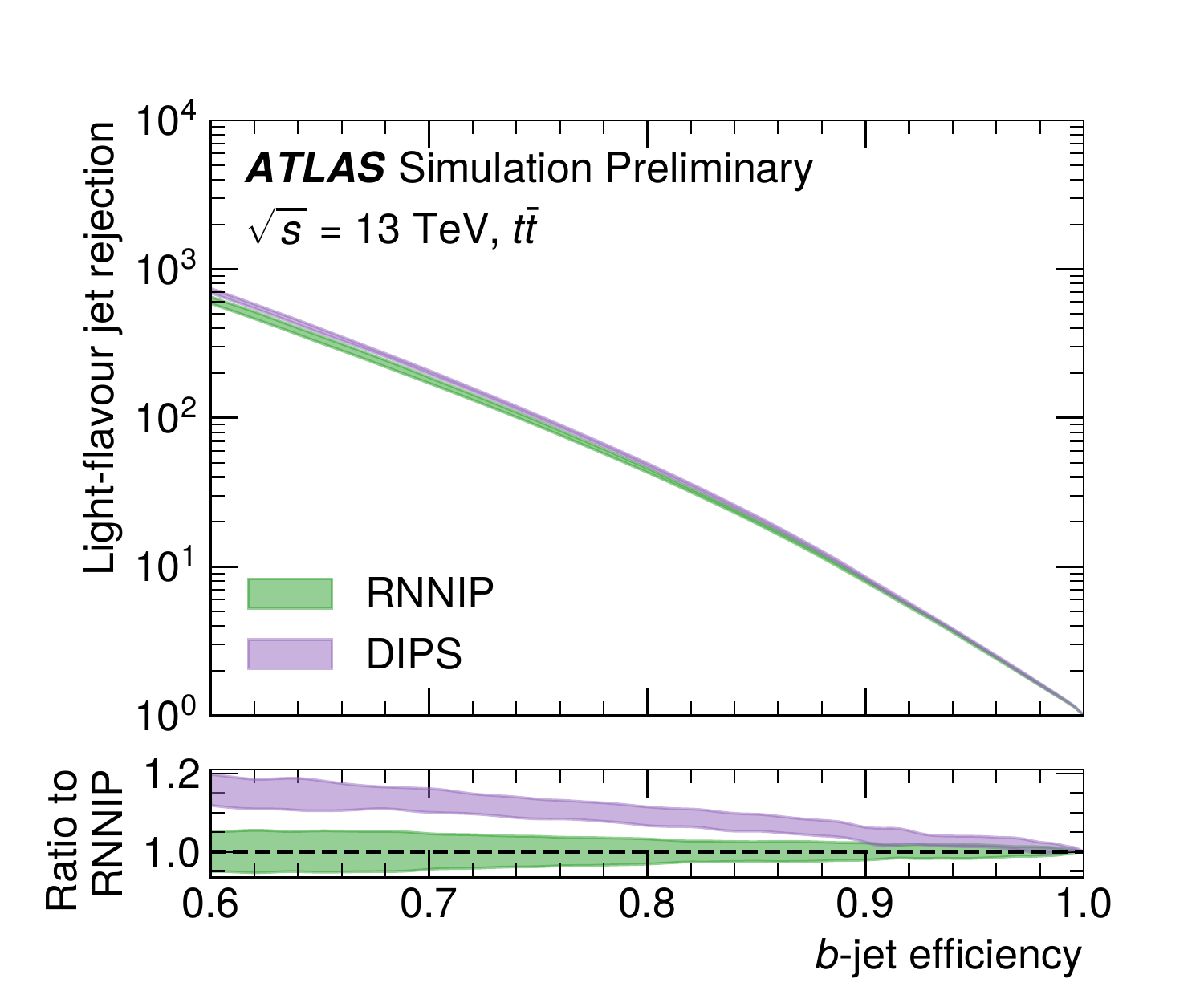}
    \caption{Performance of the RNNIP~\cite{ATLAS-RNNIP} and DIPS~\cite{ATL-PHYS-PUB-2020-014} heavy flavor identification algorithms in the ATLAS experiment. The horizontal axis shows the efficiency of correctly identifying $b$-jets, while the vertical axis shows the inverse of the efficiency of incorrectly identifying light flavor jets as $b$-jets. The violet band shows the performance the DIPS performance, while the green band represents RNNIP. 
    The width and central value of the curves represent the standard deviation and mean of the light flavor jets rejection for a given $b$-jet efficiency for five different network trainings. 
    Performances were measured for jets with a transverse momentum above 20 GeV, in a simulated dataset of top quark pairs, at center-of-mass energy of 13 TeV. }
    \label{fig:dips}
\end{figure}

A similar architecture to the Particle Flow Networks was used in the ATLAS experiment as an alternative to the RNN-based model in the context of heavy flavor identification~\cite{ATL-PHYS-PUB-2020-014}. 
The Deep Sets-based ATLAS heavy flavor discriminant (DIPS) outperforms its RNN analogous (RNNIP, described in section~\ref{sec:sequence_heavyflavor}) version of the RNNIP ATLAS heavy flavor discriminant by up to $20\%$ for $b$-jet efficiencies larger than $60\%$, while using the same inputs, as seen in Figure \ref{fig:dips}. 
DIPS has also been found to reduce significantly the training and evaluation time with respect to RNNIP, due to its paralelizability.
The ability of parallelizing computations on each sequence element is an important feature of this model, particularly for applications in which the network evaluation time is limited.

\section{Conclusion}

Sequence-based Machine Learning algorithms have a long and rich history in the context of natural language processing. 
While the idea of representing a jet as a sequence of its constituents is not new in particle physics, as evidenced by jet clustering algorithms, the usage of ML concepts exploting this representation is relatively recent when compared to computer vision algorithms. 
Even so, the application of RNNs to jet physics in particular has been a fruitful avenue of research in the past few years. 
Special attention was given to the predictive power of these models, with the successful application of LSTM-based neural network architectures to different types of jet classification tasks. 
More generally, recurrent structures were shown to be well-suited to describe jet clustering histories, leading to a full probabilistic model of a jet given its constituents. 

Recent work has also been focused on expanding the basic RNN ideas of hierarchical context learning to more physics inspired architectures, such as recursive neural networks. 
However, while these models achieve high precision when the correct choice of input structure is used - either the binary tree in a RecNN or the ordered sequence itself for RNNs - their performances can be significantly degraded given a wrong structure choice. 
This is particularly difficult to deal with when the input has variable sized length but the data structure is not obvious. 
For example, how one chooses to order the set of tracks inside the jet will depend on the task to be performed: ordering in impact parameter significance can be suited for b-jet identification, but not for quark versus gluon discrimination. 
With that in mind, algorithims in which the data structure itself is either learned, such as transformers or graphs, or invariant under certain problem transformations, such as permutation invariant sets, have shown a great potential for future studies. 

\section*{Acknowledgements}

RTdL would like to thank Michael Kagan for the help in reviewing the document; the book editors, Paolo Calafiura, David Rousseau and Kazuhiro Terao; and the other reviewers involved. This work was supported by the US Department of Energy (DOE) under grant DE-AC02-76SF00515, and by the SLAC Panofsky Fellowship.

\clearpage

\bibliographystyle{ieeetr}
\bibliography{arxiv}

\end{document}